\documentclass[sigconf]{acmart}
\usepackage{graphicx}
\usepackage{multirow}
\usepackage{booktabs} 
\usepackage{bm}
\usepackage{balance}
\usepackage{pifont}
\usepackage{enumerate}
\usepackage{algorithm}
\usepackage{algorithmic}
\usepackage{subfig}
\newcommand{\cmark}{\ding{51}}%

\hypersetup{colorlinks,urlcolor=}

\copyrightyear{2018} 
\acmYear{2018} 
\setcopyright{acmcopyright}
\acmConference{MM '18}{October 22--28, 2018}{Seoul, Korea}\acmPrice{15.00}\acmDOI{10.1145/3123266.3123312}

\begin{document}
\renewcommand{\thefootnote}{\fnsymbol{footnote}}
\title{Multimodal Co-Training \\ for Selecting Good Examples from Webly Labeled Video$^*$}

\author{Ryota Hinami}
\affiliation{\institution{The University of Tokyo}}
\email{hinami@nii.ac.jp}
\author{Junwei Liang}
\affiliation{\institution{Carnegie Mellon University}}
\email{junweil@cs.cmu.edu}
\author{Shin'ichi Satoh}
\affiliation{\institution{National Institute of Informatics}}
\email{satoh@nii.ac.jp}
\author{Alexander Hauptmann}
\affiliation{\institution{Carnegie Mellon University}}
\email{Alex@cs.cmu.edu}
\renewcommand{\shorttitle}{Multimodal Co-Training for Selecting Good Examples from Webly Labeled Videos}

\begin{abstract} 
We tackle the problem of learning concept classifiers from videos on the web 
without using manually labeled data.
Although metadata attached to videos (e.g., video titles, descriptions)
can be of help collecting training data for the target concept,
the collected data is often very noisy.
The main challenge is therefore how to select good examples 
from noisy training data.
Previous approaches firstly learn easy examples 
that are unlikely to be noise 
and then gradually learn more complex examples.
However, hard examples that are much different from easy ones 
are never learned.
In this paper, we propose an approach called multimodal co-training
(MMCo) for selecting good examples from noisy training data. 
MMCo jointly learns classifiers for multiple modalities 
that complement each other to select good examples. 
Since MMCo selects examples by consensus of multimodal classifiers, 
a hard example for one modality can still be used as a training example
by exploiting the power of the other modalities.
The algorithm is very simple and easily implemented 
but yields consistent and significant boosts in example selection and 
classification performance on the FCVID and YouTube8M benchmarks.

\end{abstract} 

\maketitle
\footnotetext[1]{This work was conducted when the first author visited Carnegie Mellon University.}
\renewcommand{\thefootnote}{\arabic{footnote}}
\setcounter{footnote}{0}

\section{Introduction} 
Learning concept classifiers for videos is a fundamental task 
for automatic understanding of video and has many practical applications.
Recent advances in deep neural networks enable
very accurate classifiers to be learned
by training on large-scale datasets with manually annotated labels 
(e.g., Kinetics dataset~\cite{Kay2017}).
However, collecting such datasets involves huge amounts of human labor.
Therefore, it is preferable to learn classifiers
without using any manually labeled data.

One of the promising approaches to learning without labeled data is 
learning from web videos containing rich metadata 
(e.g., titles, descriptions, keywords).
Such data are much easier to collect, and
datasets consisting of millions of videos are available,
e.g., YouTube8M~\cite{Abu-El-Haija2016}, YFCC100M~\cite{Thomee2016}.
Although the training data for the target concept can be 
easily collected 
by using text-based matching between the concept name and metadata,
the collected data are often noisy.
The main challenge is therefore how to select good examples from 
noisy training data.

\begin{figure}[t] 
\begin{center}
    \subfloat[Single modality]{
    \includegraphics[clip, width=1.0\columnwidth]{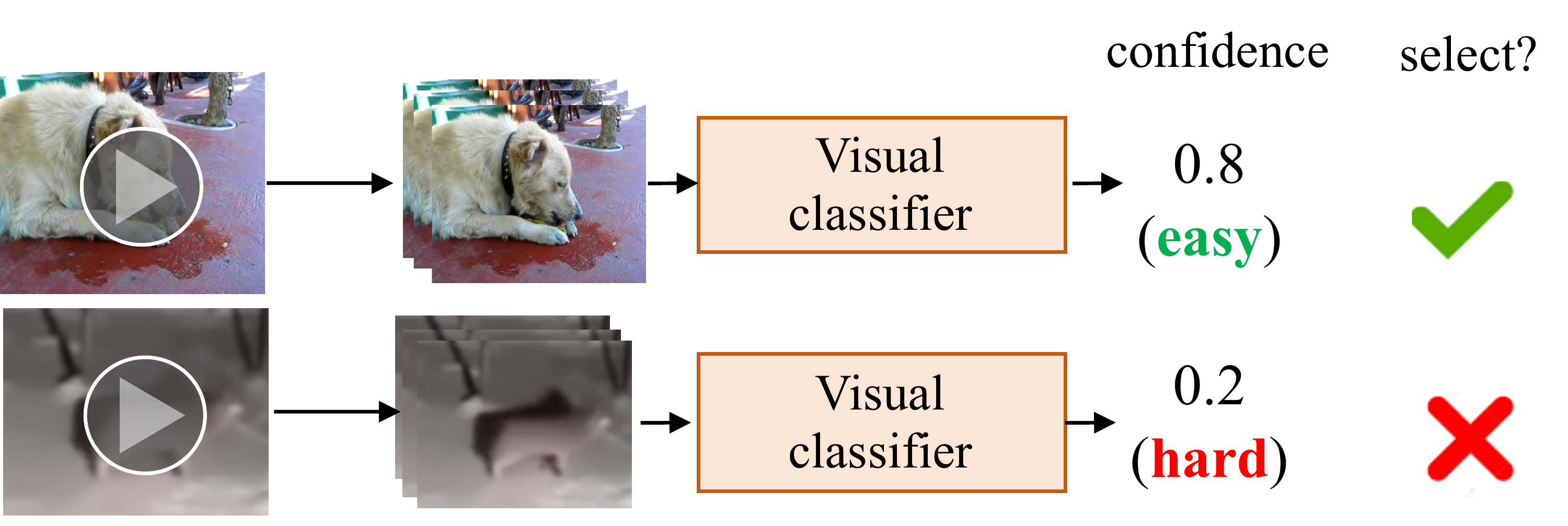}
    \label{fig:top1}
}
    \vspace{0.1mm}
    \subfloat[Multiple modalities]{
    \includegraphics[clip, width=1.0\columnwidth]{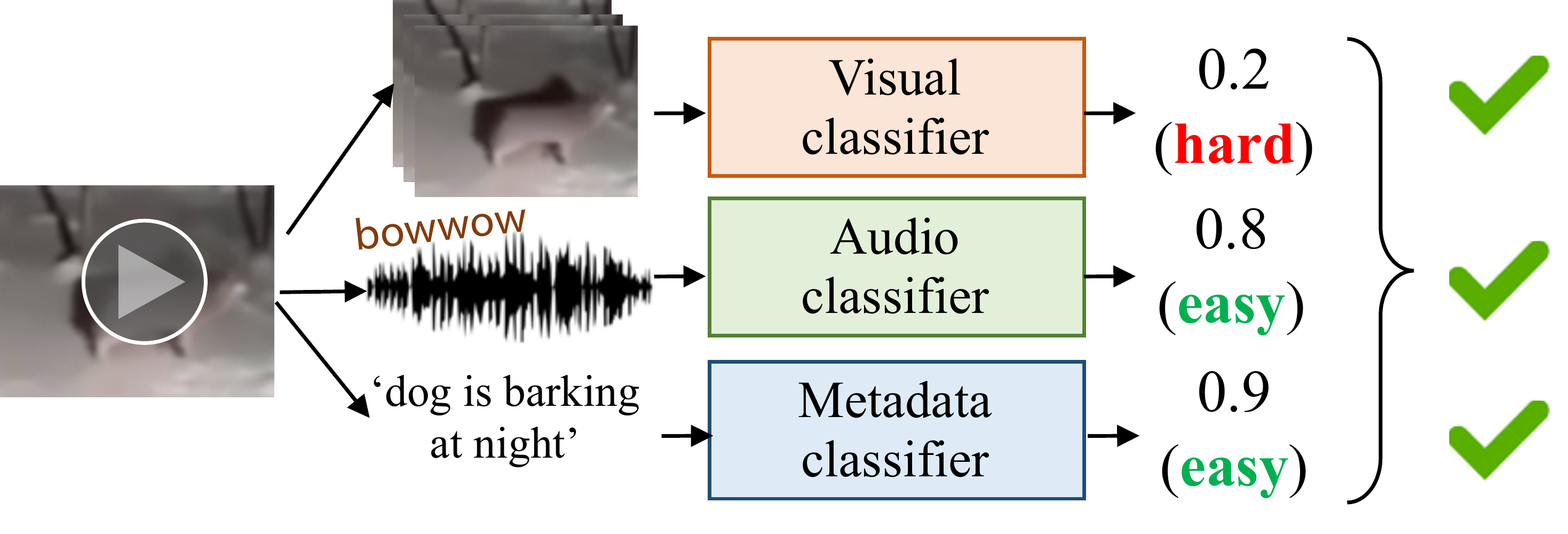}
    \label{fig:top2}
}
\vspace{-3mm}
\caption{
In webly labeled learning, confident samples are selected as 
training examples from noisy data.
(a) Previous approaches often miss hard examples
    that would potentially improve classifier performance.
(b) The proposed approach can learn hard examples by exploiting the information of other modalities.
}
\vspace{-3mm}
\label{fig:top} 
\end{center} 
\end{figure}

Liang et al.~\cite{Liang2016} proposed an approach to learning concept classifiers
from web videos with noisy labels.
The key idea is to learn easy examples 
first~\cite{Kumar2010,Bengio2009,Jiang2015}.
That is, easy examples that are very likely to be positive are learned first; 
then gradually more complex examples are learned.
This idea is similar to self-training~\cite{Yarowsky1995} 
that is widely used in semi-supervised learning.

Although this approach can select \textit{correct} examples that are easy, 
it sometimes misses \textit{hard} correct examples.
Consider hard samples that are much different from any easy sample;
a classifier trained on only easy samples will never classify 
them as positive.
In other research fields such as object detection, 
hard examples have been shown to be useful for improving a classifier.
The basic approach of hard example mining is to select examples
with high loss~\cite{Felzenszwalb2008,Girshick2014,Shrivastava2016,Schroff2015d}.
However, hard examples with high loss 
are likely to be noise in learning on noisy data.
For example, in Fig.~\ref{fig:top1}, 
while the first easy sample can be used as a positive example, 
the second hard sample cannot be used because it is possible to be noise.


How can we select good examples that are \textit{hard} and \textit{correct}?
Our key idea is to use information of multiple modalities.
Figure~\ref{fig:top} shows an example of learning a `dog' classifier.
The dog in the dark (Fig.~\ref{fig:top2}) is a hard example
with only visual information
because its appearance is much different from dogs in the light;
such an example would not be selected in self-learning based approach.
However, by using information of other modalities 
(e.g., the title `a dog is barking' or sounds of dogs barking),
we can confidently select it as a positive example.

On the basis of this intuition, 
we propose an approach called \textit{multimodal co-training} (MMCo)
for selecting good examples by using multimodal knowledge.
MMCo jointly learns the classifiers for multiple modalities and
selects examples by voting of multiple classifiers 
(e.g., taking their max score).
Therefore, a hard example for one modality can be used as a training example
by exploiting the other modalities.
The algorithm is very simple and easily implemented 
but significantly improve the performance of 
example selection from noisy training data;
precision/recall is improved from 0.66/0.64 
to 0.70/0.80 on FCVID benchmark.

In addition, we introduce an online learning algorithm for ``webly'' labeled
learning problem~\cite{Liang2016}.
Although existing approaches~\cite{Liang2016,Liang2017} uses
an alternating optimization that performs joint optimization of the model 
and the weights of the examples, 
it has trouble handling large-scale datasets consisting of millions of videos.
Here, we extend the original webly labeled learning (WELL)~\cite{Liang2016} 
to an online setting by computing the sample weights online for each minibatch.
This online WELL can handle a large-scale dataset
without degrading accuracy.
Online WELL can be extended to multiple classes easily 
and efficiently like other stochastic gradient descent(SGD)-based 
classifier learning methods.


\section{Online Webly Labeled Learning} 
Our goal is to learn a concept detector from web videos
without any manually annotated labels.
Liang et al.~\cite{Liang2016} proposed Webly Labeled Learning (WELL)
approach to this problem, where the classifier parameters and 
sample weights (selected examples) are alternately optimized. 
However, alternating optimization based on batch training
is difficult to apply to a large-scale dataset consisting of millions of videos
because all the training data should be processed in every update.
We therefore decided to extend WELL to an online setting 
by selecting samples and computing their weight for each minibatch online.
In this section, we first describe the problem of webly labeled
learning and review WELL (Sec.~\ref{sec:well}).
Then we describe online WELL (Sec.~\ref{sec:owell}).

\subsection{Background: Webly Labeled Learning}
\label{sec:well}
\textbf{Problem formulation.}
Webly labeled learning was introduced in \cite{Liang2016}.
In it, one considers the problem of learning a binary classifier $g$ 
for the concept from $N$ unlabeled web videos $\{(\mathbf{x}_n, t_n)\}_{n=1}^N$,
where $\mathbf{x}_n$ is the feature of $n$th video, 
and $t$ is the metadata (text data) attached to it.
The pseudo label of each example $y_n \in \{0,1\}$ is inferred 
by text-based matching between 
the concept name and the metadata of each video $t_i$;
i.e., $y_n$=1 is assigned if the concept name $c$ appears
in the metadata $t_n$, and $y_n$=0 if id does not appear.
The confidence $v^0_n \in \mathbb{R}$ is obtained from the accuracy 
of matching; a high confidence value is assigned 
if $c$ frequently appears in $t_n$ 
($v^0_n=1$ for negative-labeled samples).
Consequently, the problem reduces to one of learning the classifier given 
the videos with pseudo labels and their confidences 
$\{(\mathbf{x}_n, y_n, v^0_n)\}$.

The problem of webly labeled learning is related to 
semi-supervised~\cite{ChapelleOlivierandScholkopfBernhardandZien2009}
and noisy-labeled learning~\cite{Frenay2014}.
In noisy-labeled learning, a classifier is learned using samples 
with noisy (corrupted) labels ($\mathbf{x}_n$, $y_n$).
While all samples should be treated equally in noisy-labeled learning, 
in webly labeled learning,
the label confidence $v^0_n$ of each sample can be obtained from metadata 
as prior knowledge.
In semi-supervised learning, a classifier is learned 
using both labeled and unlabeled samples.  
While labeled samples are completely trusted in semi-supervised 
learning, in webly labeled learning,
samples with high confidence labels are not always positive.

\textbf{Model.}
We here describe the Webly Labeled Learning (WELL) model~\cite{Liang2016} 
that was proposed to solve the problem formulated above.
WELL is based on the self-paced curriculum learning theory proposed in \cite{Jiang2015}.
The model parameters $\mathbf{w}$ 
(e.g., linear classifier weights)
and the latent weight variable
$\mathbf{v}=[v_n, ..., v_N]^T$ are jointly learned by
alternating optimization.
The objective function is defined as
\begin{eqnarray}
\label{eq:obj}
    \min_{\mathbf{w},\mathbf{v}} E(\mathbf{w}, \mathbf{v};\lambda) = \sum_{n=1}^N v_n L(y_n, g(\mathbf{x}_n, \mathbf{w})) + f(\mathbf{v}; \lambda),
\end{eqnarray}
where $g$ is the classifier, $L$ is the loss function, 
and $\mathbf{v}=[v_1,v_2,...,v_N]^T$ denote the latent weight variables
that reflect label confidences.
Examples with larger weights tend to be learned with high priority.
$f$ is a self-paced regularizer~\cite{Kumar2010} that controls
the learning process so that easy samples with small losses tend to be 
selected earlier.
For example, the linear regularizer proposed in \cite{Jiang2015} is
\begin{eqnarray}
\label{eq:f}
    f(\mathbf{v}; \lambda) = \frac{1}{2} \lambda \sum_{n=1}^N (v_n^2 - 2v_n)
\end{eqnarray}
where $\lambda$ controls the learning pace; i.e., 
$\lambda$ is gradually increased to increase the number of samples used.
With a fixed $\mathbf{w}$, the weight variable can be optimized in 
a closed form as
\begin{eqnarray}
\label{eq:weight}
v_n = \left\{ \begin{aligned} 
    1-\tfrac{l_n}{\lambda}  \qquad& l_n < \lambda, \\
    0 \qquad& l_n \geq \lambda,
\end{aligned}\right. 
\end{eqnarray}
where $l_n$ is the loss for the $n$th sample.
Therefore, examples with losses smaller 
than $\lambda$ are used for training.

$\mathbf{v}$ is initialized by the 
pseudo label's confidence $\mathbf{v}^0$ 
computed by word matching between 
the concept name and metadata, 
which corresponds to the intuition of curriculum 
learning~\cite{Bengio2009}
that determines the learning sequence from external prior knowledge.
WELL determines the learning sequence 
by using both
prior knowledge in the manner of curriculum learning 
and the model in the manner of self-paced learning.
In this way, WELL can learn from easy examples  
as well as from more complex ones.

Below, we summarize the implementation of WELL.
\begin{enumerate}[(1)]
\setcounter{enumi}{-1}
\item{Initialize the sample weights} $\mathbf{v}$ from the label confidence $\mathbf{v}^0$.
\item{Update the model} $\mathbf{w}$ with fixed sample weights $\mathbf{v}$.
\item{Compute the losses} of all examples by using the updated model.
\item{Compute the weights} $\mathbf{v}$ of all examples by using Eq.~\ref{eq:weight}.
\item{Increase age $\lambda$.} \qquad 
    Repeat (2)--(5) until convergence.
\end{enumerate}

\subsection{Online Webly Labeled Learning}
\label{sec:owell}
Although WELL learns the classifier by alternating minimization,
it is not appropriate for the training on a large-scale dataset.
In alternating optimization, all training samples are used to 
optimize the model $\mathbf{w}$ in every iteration, 
wherein the sample weights $\mathbf{v}$ are fixed when optimizing the model.
If the training data is very large (e.g., millions of samples),
it takes a very long time to complete one iteration,
which makes convergence very slow.
In addition, it is generally infeasible to put whole dataset in memory. 
Hence, we decided to extend the webly labeled learning model 
in \cite{Liang2016} to the online learning based 
on stochastic gradient decent (SGD).

The main problem is how to select optimal samples and 
compute their weights in a minibatch.
If the sample weights $\mathbf{v}$ are given, the model can be 
stochastically optimized over the minibatch in the manner of SGD.
Although alternating optimization generally requires all samples 
in order to optimize each variable, 
in the binary scheme of WELL, 
the weight $v_n$ can be computed using only the loss of the sample $l_n$
and the model age $\lambda$ using Eq.~\ref{eq:weight}.
Therefore, we can compute the sample weight $v_n$ online
after computing the loss of each sample $l_n$,
where the model is updated with the weighted loss $v_nl_n$.
The age is determined by the scheduler described below.

More specifically, the online webly labeled learning algorithm is as follows.
\begin{enumerate}[(1)]
\setcounter{enumi}{-1}
\item{Initialize model $\mathbf{w}$ and age $\lambda$.}
\item{Randomly sample a minibatch with $N_m$ examples.}
\item{Compute the losses} $l_1, .., l_{N_m}$ using the model $\mathbf{w}$. 
    Cross entropy loss is used in this study.
\item{Compute the weights} $v_1, .., v_{N_m}$ of the samples in the minibatch 
    with Eq.~\ref{eq:weight} and the computed losses 
    (the confidences of the pseudo labels $\mathbf{v}^0$ 
    are used in the first few iterations.)
\item{Update} $\mathbf{w}$ by using the weighted losses, i.e.,
    $\sum_{i=1}^{N_m} v_il_i$
\item{Increase the age $\lambda$} based on the age scheduler described below.
\end{enumerate}
Repeat (2)--(5) until convergence. 

We should note that the training cost of online WELL is less than that of WELL.
WELL needs to compute the loss twice 
in each iteration in order to update the weights $\mathbf{w}$ and the model $\mathbf{v}$.
On the other hand, we have to compute the loss only once;
sample weights $\mathbf{w}$ are computed using the loss,
and the model $\mathbf{v}$ is updated using the weighted loss.

\textbf{Age scheduling.}
Model age $\lambda$ is an important parameter that determines
which samples to select as examples.
Manually determining the age schedule 
(initial value, early stopping value, etc.)
is hard; in practice, 
the age is controlled in accordance with the rate of used samples in the 
positive pool $p$ (e.g., 30\% positive samples ($p$=0.3) are first used,
and $\lambda$ is gradually increased until 60\% 
positive samples ($p$=0.6) are used), which can be specified more intuitively.
The age $\lambda$ is computed from $p$ by sorting all positive samples,
where the loss at the $100p$th percentile is $\lambda$.
However, in online learning, not all of the training samples are available 
for the updates, the age cannot be computed from $p$. 

To control $\lambda$ in online learning, 
we introduce a FIFO queue that stores the losses of the previous iterations.
We can compute $\lambda$ from $p$ by regarding the samples 
in the queue as a whole training set.
There are two approximations here.
First, the samples in the queue approximate the whole set of training samples;
the size of the queue should be large enough to hold enough samples.
Second, the losses in the queue are computed from the previous model;
they are different from the loss computed by the current model.
The size of the queue should not be too large because of this.
Although there is a trade-off between these two approximations,
the same performance as in batch training can be achieved
by selecting an appropriate queue size.

\textbf{Multiple classes.}
Online WELL is easily extended to multiple classes 
like other multilabel classification models learned by SGD.
Given a dataset $\{(\mathbf{x}_n,\mathbf{y}_n)\}_{n=1}^{N}$, 
the goal is to learn a multilabel classifier that 
outputs the confidence of $C$ classes, 
where $\mathbf{y}_n = [y_{n,1}, ..., y_{n,C}]^T\in \{0,1\}^{C}$ 
are labels for $C$ classes. 
Note that we do not assume that the classes are mutually exclusive;
therefore, the samples can be positive for multiple classes.
For each iteration,
the age $\lambda_{C}$ is determined for each class independently
following the age scheduling described above,
whereby the weights of the samples $\mathbf{V}\in \mathbb{R}^{N \times C}$ 
are computed for every class. 
The loss in a minibatch can be computed as 
\begin{eqnarray}
\label{eq:multiclass}
    \sum_{n=1}^N \sum_{c=1}^{C} \mathbf{V}_{n,c} L(\mathbf{y}_{n,c}, g(\mathbf{x}_n, \mathbf{w}, c)),
\end{eqnarray}
where $g(\cdot, \cdot, c)$ is the prediction for the $c$th class.
By sharing the computation, this algorithm enables multiple classifiers 
to be learned much more efficiently than learning each classifier independently,
especially computationally heavy classifiers like deep neural networks.


\begin{figure*}[t] 
\begin{center}
\includegraphics[width=1.00\linewidth]{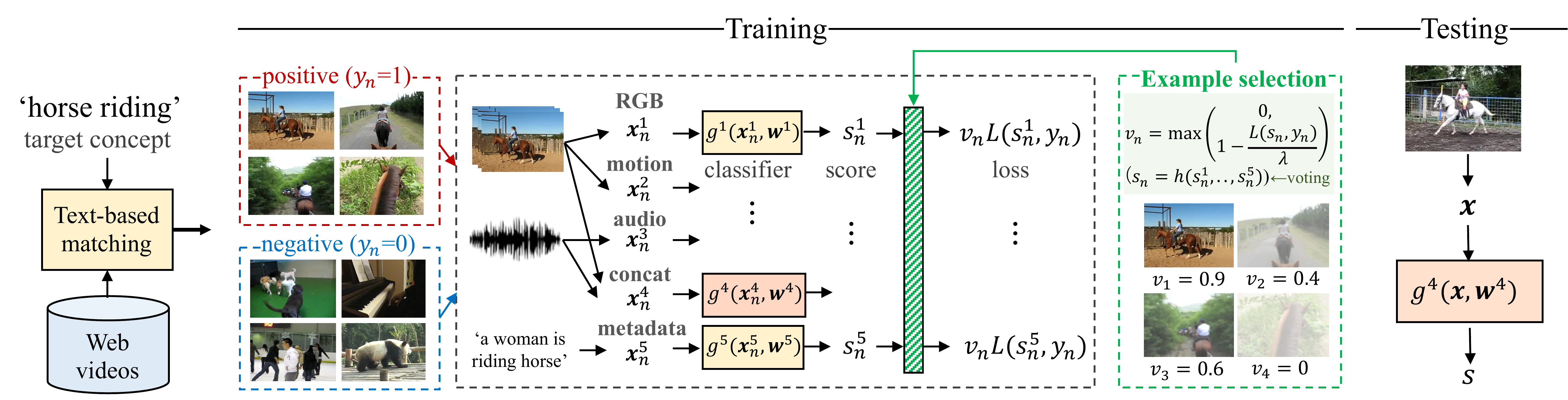}
\vspace{-6mm}
\caption{Overview of proposed approach using MMCo. 
Classifiers of multiple modalities are jointly learned.
While the loss is computed for each classifier independently, 
weights of examples are determined by voting of multiple classifiers.
    }
\vspace{-4mm}
\label{fig:overview} 
\end{center} 
\end{figure*}

\section{Multimodal Co-Training} 
\label{sec:mmco}
In webly labeled learning, easy examples with small losses 
are preferentially used as training data.
Correct but hard examples that would potentially improve
performance are never learned.
As a solution to this problem, we propose multimodal co-training (MMCo)
as a way to use multimodal information in web videos  
(e.g., visual, audio, and metadata), which enables us to exploit hard examples.
For example, a sample with a small loss in the audio modality 
but large loss in the visual modality could be a hard (but correct) 
sample from the visual perspective (a dog at night maybe).
MMCo can utilize such samples by selecting examples 
on the basis of the consensus of the classifiers of individual modalities.
Hard examples for one modality can still be used as a training example
by exploiting the power of the other modalities.
In this way, the classifiers of individual modalities mutually 
enhance each other.
The underlying motivation of our approach is very close to that of
co-training in semi-supervised learning~\cite{Blum1998}, 
which is discussed in Sec.~\ref{sec:rel}.

\subsection{Model}
We jointly learn multiple modality classifiers.
Given a dataset consisting of multiple modality features and 
pseudo labels $\{(\mathbf{x}_n^1,..., \mathbf{x}_n^M), y_n\}_{n=1}^N$,
the goal is to learn classifier parameters for 
multiple modalities $(\mathbf{w}^1, ..., \mathbf{w}^M)$, 
where $M$ is the number of modalities.
The classifier of each modality $g^m(\mathbf{x}_n^m, \mathbf{w}^m)$
is learned with the loss computed in the 
same way as single modality (Eq.~\ref{eq:multiclass}).
The only difference is the way of computing the weight variable 
$v_n$.
$v_n$ reflects the label confidence, and thus
does not depend on modality $m$, i.e.,
the same weight is used across all modalities.
Therefore, we determine the selected samples and their weights
by the consensus of classifiers of all modalities.

{\bf Sample selection.}
We here describe how to select examples in MMCo,
i.e., how to compute sample weights $v_n$.
In a previous approach, the weight is computed from the loss using Eq.~\ref{eq:weight},  
in which the loss is computed using the cross entropy between 
the confidence $g(\mathbf{x},\mathbf{w})$ and label $y$.
The confidence is computed by the voting of multiple classifiers.
Specifically, the sample weight $v_n$ is calculated as follows:
\begin{eqnarray}
\label{eq:aaa}
    v_n = \mathrm{max}(0, 1-\frac{L(s_n, y_n)}{\lambda}),
\end{eqnarray}
where the confidence $s_n$ of the sample is 
\begin{eqnarray}
s_n = h(g^1(\mathbf{x}_n^1, \mathbf{w}^1), ..., g^M(\mathbf{x}_n^M,\mathbf{w}^M)), 
\end{eqnarray}
where $h$ is the voting function that takes the input of multiple classifier confidences $\{g^m(\mathbf{x}_n^m, \mathbf{w}^m)\}_{m=1}^{M}$.
We introduce three types of simple voting scheme, namely, 
\textit{max} ($\max_{m} g^m(\mathbf{x}_n^m, \mathbf{w}^m)$), 
\textit{average} ($\frac{1}{M}\sum_{m} g^m(\mathbf{x}_n^m, \mathbf{w}^m)$), 
and \textit{product} ($\prod_{m} g^m(\mathbf{x}_n^m, \mathbf{w}^m)$).
This enable us to use hard examples of each classifier.
For example, with \textit{max} voting, 
if $g^1(\mathbf{x}_n^1, \mathbf{w}^1)$ is small and 
$g^2(\mathbf{x}_n^2, \mathbf{w}^2)$ is large,
a large weight is assigned to this sample 
although it is a hard example for $g^2$.
Our discussion for a single class can be extended to multiple classes
in the same manner of Sec.~\ref{sec:owell}.

\subsection{System Design}
The modalities used in this study are RGB, motion, audio, and metadata
(see Sec.~\ref{sec:exp_set} for the details of the features).
In addition, the concatenated features of all modalities 
extracted from video (RGB, motion, and audio) are 
used as another modality, which we call \textit{concat} feature.
In training, the classifiers of all modalities are used and 
mutually enhanced by our co-training approach.
In testing, only the classifier of the concat feature is used; 
metadata information is not used, because we assume that metadata 
is not available for the test videos
(It is also possible to use a metadata classifier 
for classifying the videos with metadata).
We found that using the classifier for concatenated features 
worked better than using classifiers for multiple modalities.

The important point here is that a modality that cannot be 
used in testing can be used for training. 
For example, 
we can train the classifiers for the metadata modalities even 
if metadata is not available in the testing;
a video-content classifier can be improved by co-training with 
a metadata classifier. 
On the other hand, the original WELL is based on self-paced learning,
and thus, it cannot fully exploit metadata modality information 
(it can only be used to extract pseudo labels).

Although we can use any type of classifier for each modality,
from the aspect of large-scale retrieval, 
it is better to use a linear classifier in testing 
than a more complex classifier such as a deep neural network.
Large-scale retrieval is one of the main applications of video concept classifier learning.
To retrieve from large amounts of video, the classifier 
should be scalable.
A linear classifier is easy to scale for a large database,
whereas a deep neural network is difficult to scale.
Although the scalable classifier should be used in testing,
any type of classifier can be used for another modality in training.
Therefore, our approach can exploit the power of a strong classifier 
(e.g., a deep neural networks)
to select good examples from noisy training data,
while previous self-paced learning-based approaches select examples 
using only the testing classifier.

\subsection{Relationship with prior work} 
\label{sec:rel}

\textbf{Self-training, co-training:}
Self-training~\cite{Yarowsky1995} and co-training~\cite{Blum1998} 
are commonly used techniques in semi-supervised learning. 
In self-training,
a classifier is first learned with a small amount of labeled data; 
then, it is applied to unlabeled data, wherein
the most confident samples are added as new training examples.
On the other hand, 
co-training iteratively train classifiers on two different views,
where new training examples of each classifier are determined by 
the prediction of the other classifier. 
While the idea of WELL~\cite{Liang2016} is based on self-training 
that selects examples for the classifier by using its own predictions,
MMCo incorporates the idea of co-training; 
examples for each classifier are selected 
on the basis of the predictions of other classifiers 
as well as its own predictions.
Unlike co-training in semi-supervised learning,
MMCo does not involve an iterative process, 
so it can easily handle more than two classifiers.


\textbf{Multimodal ensemble learning:}
Our approach selects examples on the basis of the consensus of classifiers
for multimodal features, 
which is the same idea as ensemble learning~\cite{Dietterich2002a}
that makes predictions by combining multiple classifiers.
It is known that an effective ensemble learning system 
should consist of diverse classifiers~\cite{SUN2010119,Sun2007}, 
and several studies have investigated the use of complementary multimodal 
information in videos for ensemble learning~\cite{Hsu2004,Chang2007,Fan2007}.
While ensemble learning is generally used to improve prediction accuracy,
our approach uses an ensemble of multimodal classifiers to select 
good examples from noisy training data.
Although we investigated only simple voting of classifiers, 
other ensemble learning techniques~\cite{Freund1995,Wolpert1992,Breiman1996} 
could be integrated into our approach.


\textbf{Learning from search engine results:}
While we tackle webly labeled learning~\cite{Liang2016,Liang2017} 
that learns from web videos with weak annotations,
there is another type of webly supervised classifier learning 
which directly uses results returned by 
search engines~\cite{Fergus2005,Chen2015a,Yeung2017,Chatfield2012,Chatfield2015}.
This type typically uses a search engine (e.g., YouTube) to search by concept name 
and learns a classifier by using the retrieved results as positive samples.
Since the challenge is similar to our problem, i.e., 
how to handle noise in the search results,
our approach could be applied for this problem by 
using the search results as positive-labeled noisy training data
and the search rank to compute the initial label confidence 
(i.e., top-ranked videos are likely to be positive).



\begin{figure*}[t]
    \begin{center}
	\subfloat[]{
		\includegraphics[clip, width=0.66\columnwidth]{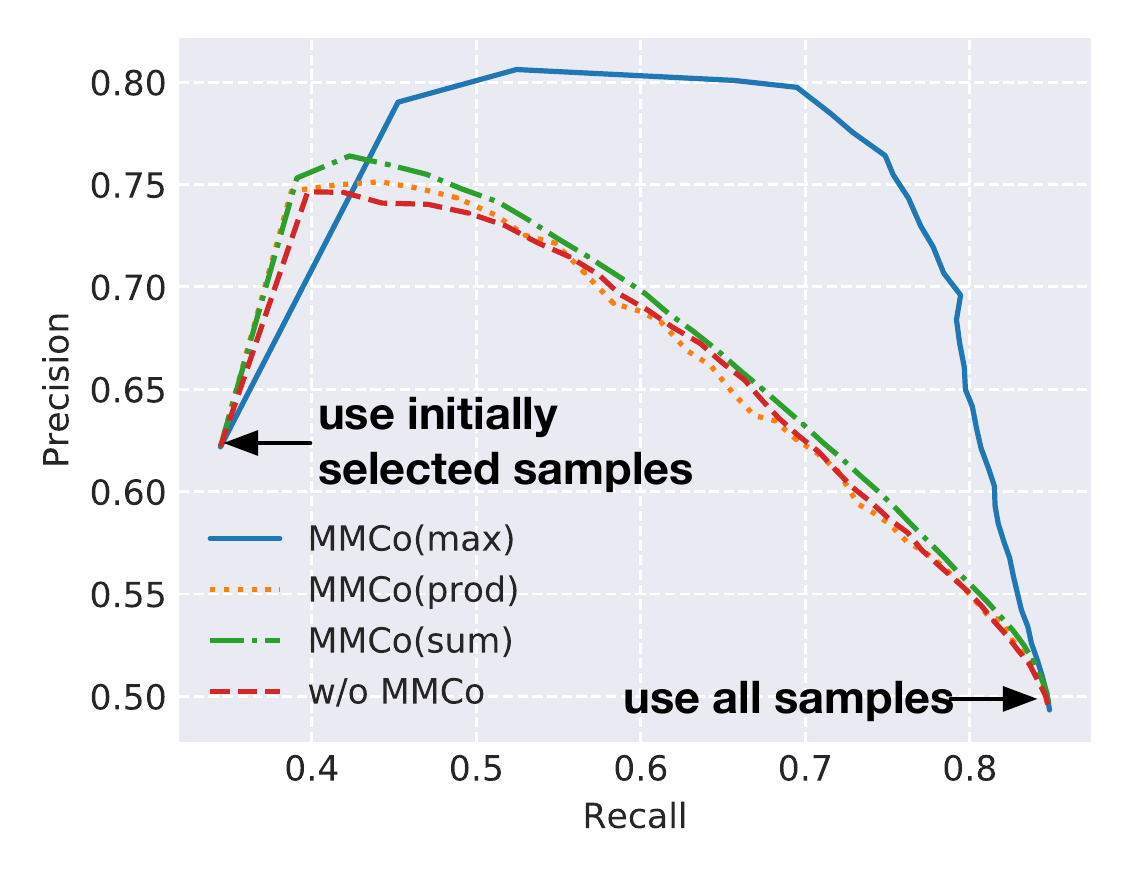}
        \label{fig:pr1}
    }
	\subfloat[]{
		\includegraphics[clip, width=0.66\columnwidth]{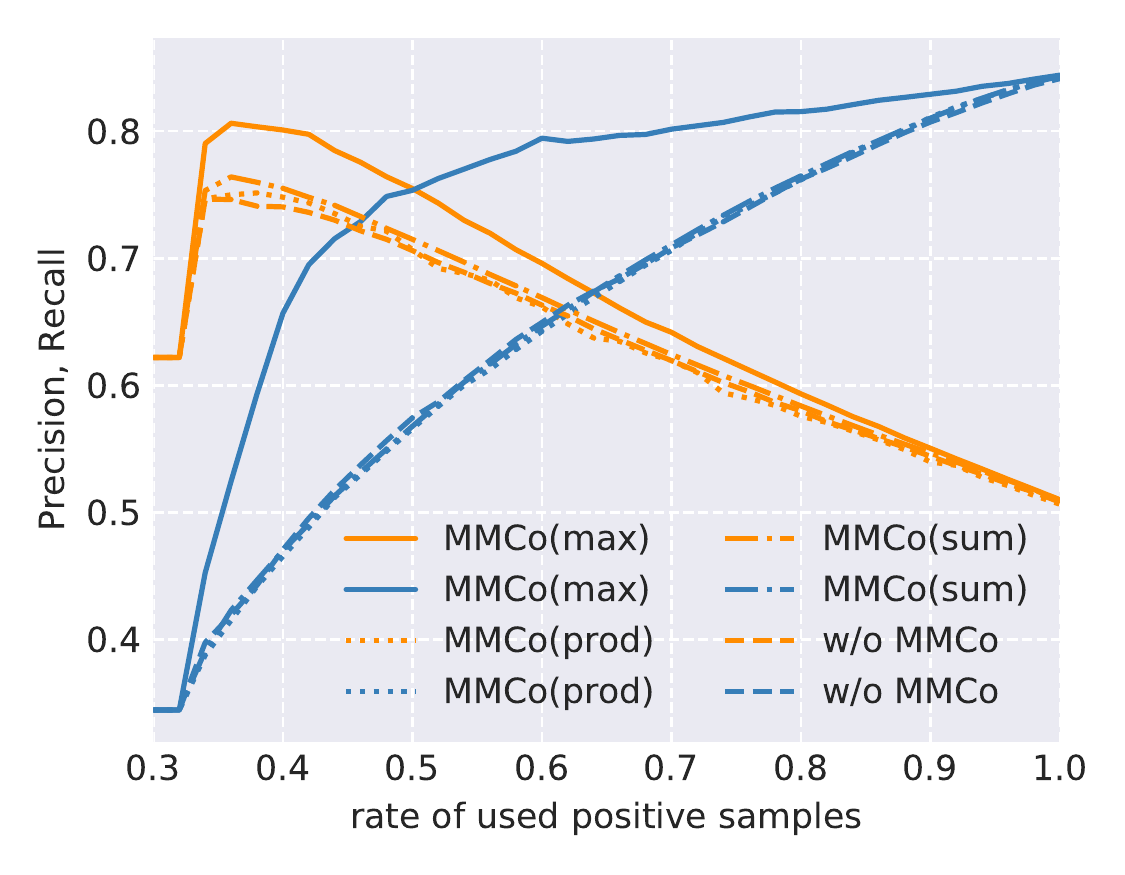}
        \label{fig:pr2}
    }
	\subfloat[]{
		\includegraphics[clip, width=0.66\columnwidth]{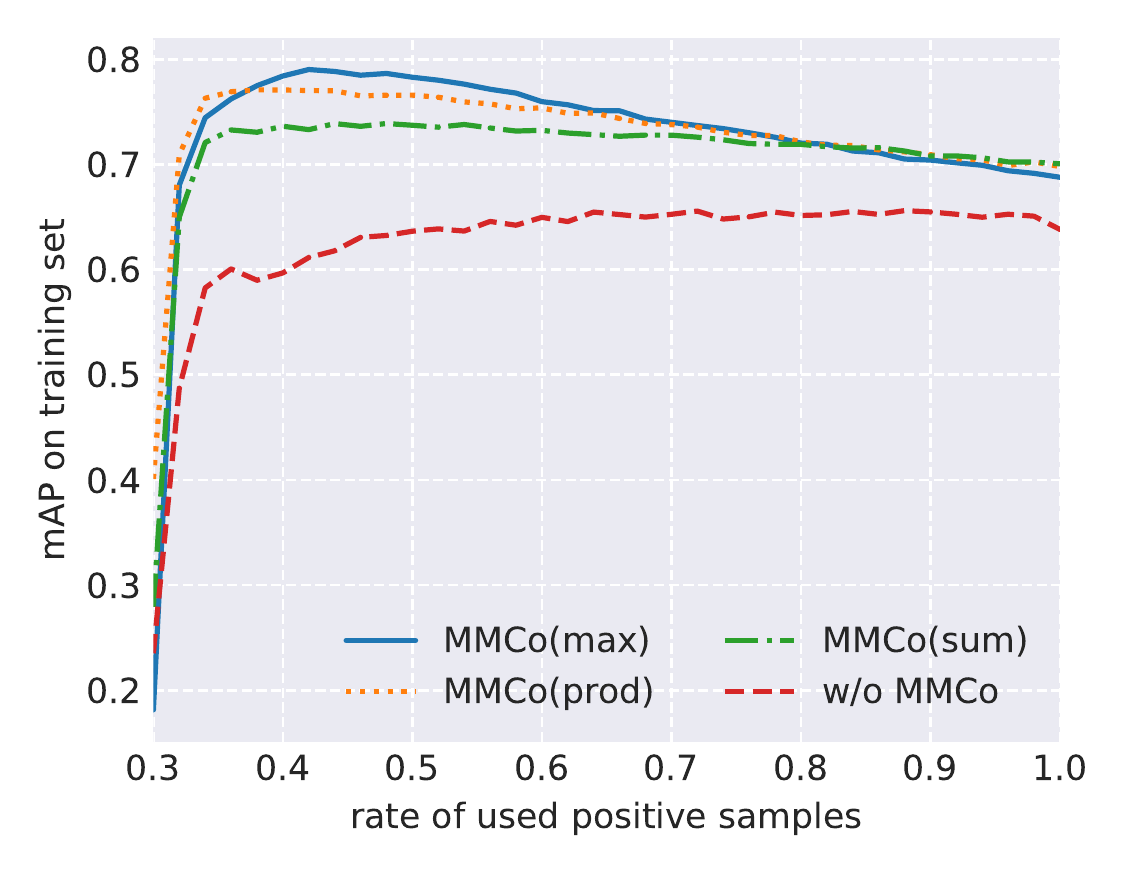}
        \label{fig:pr3}
    }
    \vspace{-4mm}
    \caption{Sample selection quality on FCVID training set. 
    (a) precision/recall curve. 
        (b) transition of \textcolor[rgb]{0.23,0.49,0.71}{recall (blue)} 
        and \textcolor[rgb]{0.99,0.55,0.14}{precision (orange)}. 
        (c) transition of mAP on training set}
    \vspace{-2mm}
    \label{fig:pr}
    \end{center}
\end{figure*}

\section{Experiments} 
\subsection{Experimental Settings}
\label{sec:exp_set}
{\bf Datasets.}
The main evaluation used Fudan-columbia Video Dataset (FCVID)~\cite{Jiang2018}
as the main evaluation.
FCVID contains 91,223 videos from YouTube, totaling 4232 hours of videos.
The dataset contains ground truth annotations at the video level
for 239 categories including activities, objects, scenes, etc.
We used the standard partitioning of the training/test set 
provided by \cite{Jiang2018}.
We did not use the manual labels in training.
Instead, we inferred pseudo labels from the metadata (title and description)
in the same manner as \cite{Liang2016}.
In addition, we used the YouTube8M dataset~\cite{Abu-El-Haija2016} 
in the larger scale evaluation.

{\bf Features.}
We first extracted features from videos and 
used them as classifier inputs.
In the evaluation of FCVID,
we used the following features for each modality.
{\bf RGB:} Frame-level features were first extracted from keyframes by 
VGG~\cite{Simonyan2015} trained on ILSVRC,
and a video-level feature was then computed by average pooling;
these were the same features used in \cite{Liang2016}.
{\bf Motion:} P3D Resnet features were extracted in the manner described
in \cite{Qiu2017}.
We used the model trained on the Kinetics dataset~\cite{Kay2017},
which is provided by \cite{Hara2017}.\footnote{https://github.com/kenshohara/3D-ResNets-PyTorch}
{\bf Audio:} We extracted 128-dimensional VGGish features produced 
from a VGG-like audio classification model~\cite{Hershey2017}.
We used the model trained on the YouTube8M dataset~\cite{Abu-El-Haija2016} 
provided by Google.\footnote{https://github.com/tensorflow/models/tree/master/research/audioset}
{\bf Metadata:} TextCNN features were extracted from the title
of the YouTube video following \cite{Wang2017}.
We used the model provided by \cite{Wang2017}, 
which was trained on the YouTube8M dataset.\footnote{https://github.com/hrx2010/YouTube8m-Text}
{\bf Concat:} The features of RGB, Motion, and Audio are concatenated. 

In the YouTube8M evaluation, we used the officially provided
RGB and audio features (1024-dimensional GoogLeNet~\cite{Szegedy2015a} 
feature and 128-dimensional VGGish feature).
For metadata, we used the same features as FCVID evaluation.

{\bf Classifiers.}
We used a linear classifier for all modalities except audio.
A multi-layer perceptron with one hidden layer (4096 units) was used
for audio feature classifier.

{\bf Training.}
The model age $\lambda$ was determined from the rate of used 
positive examples $p$;
$p$ was initialized as 0.3, which is increased by 0.05 every five epochs 
and early stopped at $p$=0.6.
The training labels (i.e., pseudo labels $y_n$ and their confidences
$v_n^0$) were obtained by word hard matching 
in the same way as \cite{Liang2016}.
Adam optimizer~\cite{Kingma2015} was used to update the model
with initial learning rate $10^{-3}$.

\begin{table}[t] 
\begin{center}
\begin{tabular}{@{}l c c c c c@{}}
\toprule
     & \multicolumn{2}{c}{\bf Modality} & & \\\cmidrule(lr){2-3}
    Method &  Training & Testing & Voting &   mAP   \\\midrule
    BatchTrain~\cite{Liang2016}          & VGG& VGG & - & 0.469 \\
    SPL~\cite{Kumar2010}                 & VGG& VGG & - & 0.414 \\
    GoogleHNM~\cite{Toderici}            & VGG& VGG & - & 0.472 \\
    BabyLearning~\cite{Liang}            & VGG& VGG & - & 0.496 \\
    WELL~\cite{Liang2016}                & VGG& VGG & - & 0.565 \\
    WELL-MM~\cite{Liang2017}             & VGG+MM& VGG & - & 0.615 \\
    WELL*                                & Concat& Concat & - & 0.619 \\\midrule
\textbf{Online WELL}                              & Concat& Concat & - & 0.621 \\\midrule
    \multirow{3}{*}{\shortstack{\textbf{Online WELL} \\ \textbf{w/ MMCo}} }
    & \multirow{3}{*}{\shortstack{VGG + P3D \\ +Audio \\ +TextCNN \\ +Concat} }
      &\multirow{3}{*}{Concat}& Max            & \textbf{0.663} \\
    & & & Sum            & 0.622 \\
    & & & Product        & 0.630 \\
\bottomrule
\end{tabular}
\end{center}
\vspace{0mm}
\caption{Performance comparison on FCVID test set}
\vspace{-6mm}
\label{tab:sota} 
\end{table}

\subsection{Comparison with Baselines}
We compared final classification performances of the classifiers
trained on the web videos on FCVID.
The main objective of this comparison was to determine whether
online WELL achieves comparable performance to the conventional WELL 
based on alternating optimization
and whether MMCo improves performance.
To this end, we reproduced WELL with the same features and parameters
and compared it with the online WELL with and without MMCo.
We used concat features in our reproduction of WELL for a fair comparison.

Table~\ref{tab:sota} shows the results.
Online WELL achieves comparable performance 
to WELL using the same features.
MMCo with max voting improves mAP by 4.2 points 
relative to WELL without MMCo.
It also significantly outperforms the other baselines~\cite{Kumar2010,Toderici,Liang}.
Although the different features are used for the comparison, 
our approach improves even on WELL that is much better than the 
others with the same VGG features.
In addition, WELL-MM~\cite{Liang2017} and MMCo are complementary extensions of WELL.
WELL-MM infers pseudo labels and their confidences using
multimodal knowledge instead of word matching, 
which can easily be incorporated into our approach 
simply by changing pseudo labels.

\subsection{Evaluation of Example Selection}
\label{sec:select}
To show that MMCo selects good examples from noisy training data, 
we evaluated the performance of example selection.
In webly labeled learning, the samples with the loss smaller than the age
were used as positive samples as described in Eq.\ref{eq:weight};
we measured the quality of selected positive samples 
by recall, precision, and AP.
Recall and precision were computed as the number of selected positive 
samples divided by the total number of positive samples 
and the total number of selected samples, respectively, and
AP was computed by the ranking of losses for all positive samples.
We used online WELL and compared with and without MMCo.
Voting schemes were also compared.
The baseline (w/o MMCo) computed the sample weights 
using one classifier 
for concat features while the proposed approach (w/ MMCo)
computed weights by the voting of classifiers of all five modalities.

Figure~\ref{fig:pr} shows the precision/recall curve and performance during the training.
The proposed MMCo with max voting ({\it MMCo (max)})
performed significantly better than the baseline.
When the rate of used samples was 0.6 (in Fig.~\ref{fig:pr2}),
the precision/recall of MMCo (max) was \textbf{0.70/0.80}, which is much better than 
the results of the baseline without MMCo (0.66/0.64) and 
the results using all samples with pseudo labels (0.49/0.85).
These results clearly demonstrate the effectiveness of MMCo.
The experiments in Sec.~\ref{sec:hard} provide a more detailed analysis 
of example selection and discuss why max voting performs 
the best.

\subsection{Large-Scale Evaluation on YouTube-8M}
\begin{figure}[t] 
\begin{center}
\includegraphics[width=1.00\linewidth]{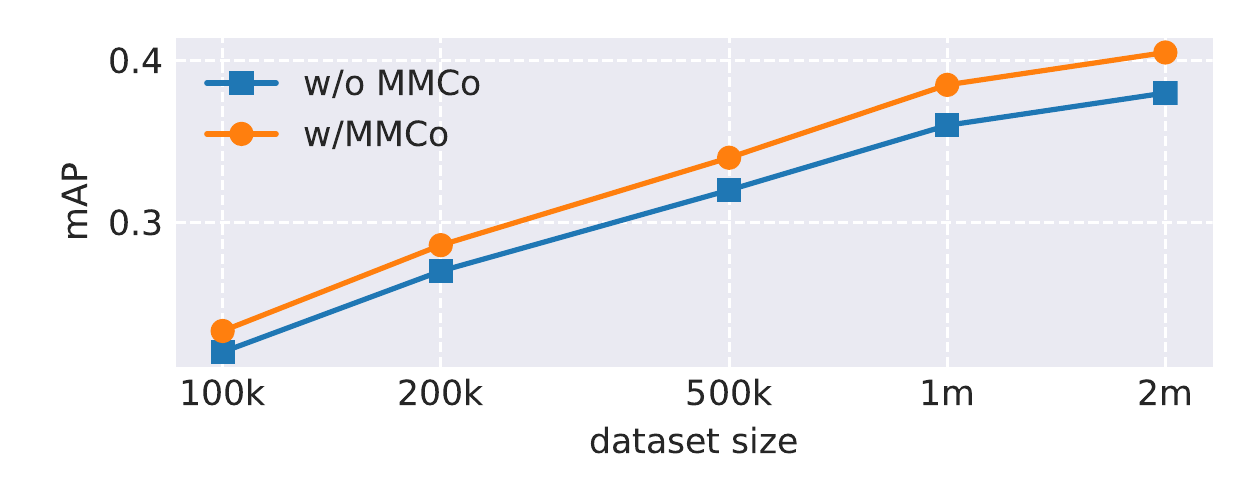}
\vspace{-5mm}
\caption{Dataset size vs test mAP on YouTube8M.}
\vspace{-3mm}
\label{fig:size}
\end{center} 
\end{figure}

The YouTube-8M dataset contains millions of videos; 
it is too big to be handled 
by the conventional approach based on alternating minimization.
We conducted an evaluation on the YouTube-8M dataset to determine 
whether the performance of online WELL gets better 
on larger-scale datasets.
We varied the size of the dataset by randomly selecting samples.
We evaluated mAP for the 500 most frequently appearing classes from all 4716 classes,
because samples from many classes in YouTube8M rarely appear 
when the size of the dataset is small.
Figure~\ref{fig:size} shows the performance of our approach with and without 
MMCo (max voting) for different subset sizes.
mAP increases along with the size of the noisy training data up to two million samples.
MMCo consistently improves mAP for any dataset size 
(about a 5\% relative improvement).
The results show that online WELL has the potential to increase its accuracy, 
because it can handle an unlimited number of videos and noisy training data is easy to collect.

\section{Detailed Analysis}
We conducted a more detailed analysis 
to better understand the behavior of MMCo.
We used max voting unless otherwise specified.

\begin{table}[t] 
\begin{center}
\begin{tabular}{@{}l c c c c c c@{}}
\toprule
             & \multicolumn{3}{c}{\bf mAP} & \multicolumn{3}{c}{\bf Recall}  \\ 
 \cmidrule(lr){2-4}\cmidrule(lr){5-7}
             & easy & normal   & hard     & easy     & normal      & hard \\
\midrule
w/ MMCo& 0.655 & \textbf{0.642} & \textbf{0.506} & 0.823 & \textbf{0.709} & \textbf{0.579} \\
    w/o MMCo    & \textbf{0.700} & 0.592 & 0.256 & \textbf{0.857} & 0.648 & 0.299 \\
\bottomrule
\end{tabular}
\end{center}
    \vspace{0mm}
    \caption{Sample selection performance in terms of the difficulty (easiness) of samples. 
    The same number of positive-labeled samples is used for both conditions ($p$=0.6).}
    \vspace{-5mm}
\label{tab:hard1} 
\end{table}

\begin{figure}[t] 
\begin{center}
\subfloat[Without MMCo]{
    \includegraphics[clip, width=0.5\columnwidth]{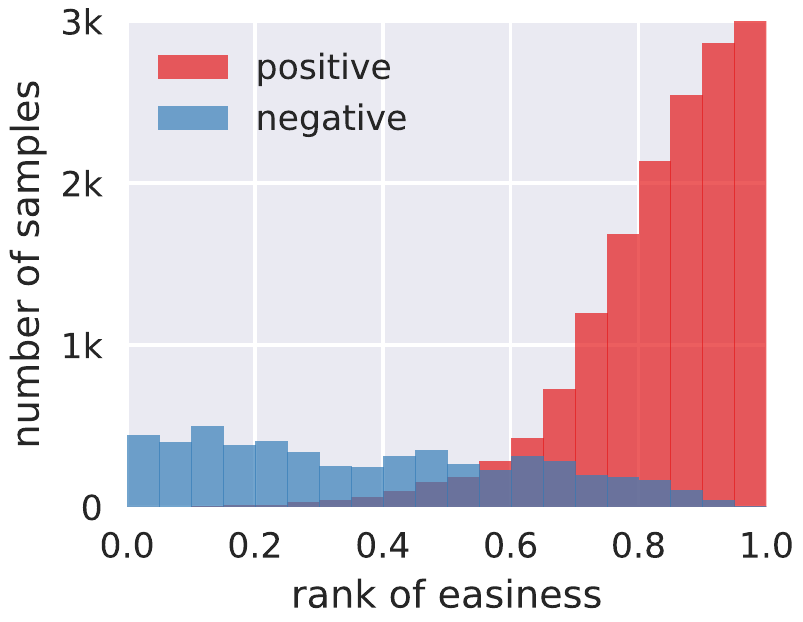}
    \label{fig:dist1}
}
\subfloat[With MMCo]{
    \includegraphics[clip, width=0.5\columnwidth]{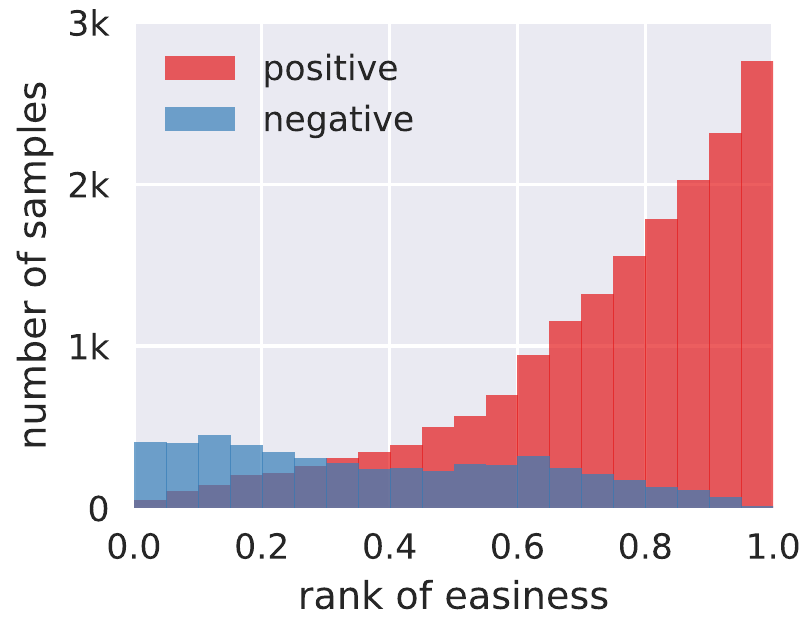}
    \label{fig:dist2}
}
\vspace{-3mm}
\caption{Easiness distribution of selected examples}
\vspace{-3mm}
\label{fig:hard2} 
\end{center} 
\end{figure}

\begin{figure*}[t] 
\begin{center}
\subfloat[]{
    \includegraphics[clip, width=0.57\columnwidth]{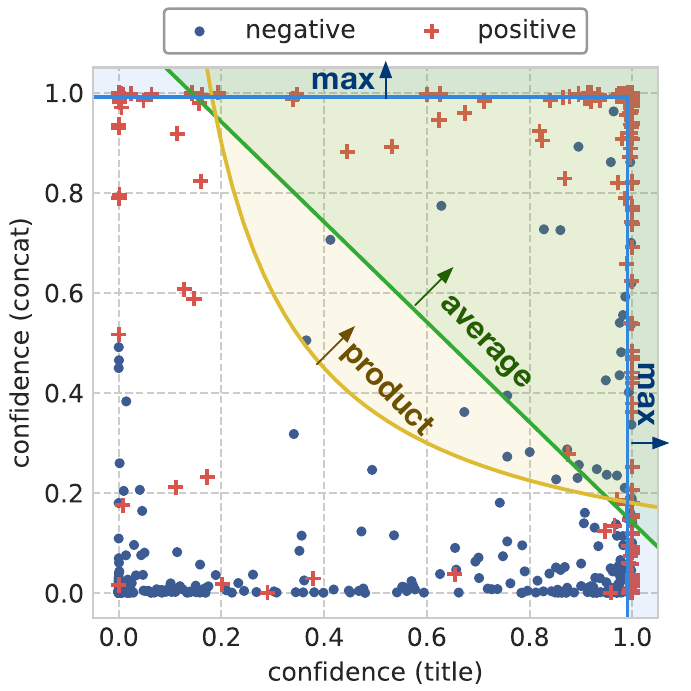}
    \label{fig:hard3_1}
}
\subfloat[]{
    \includegraphics[clip, width=1.43\columnwidth]{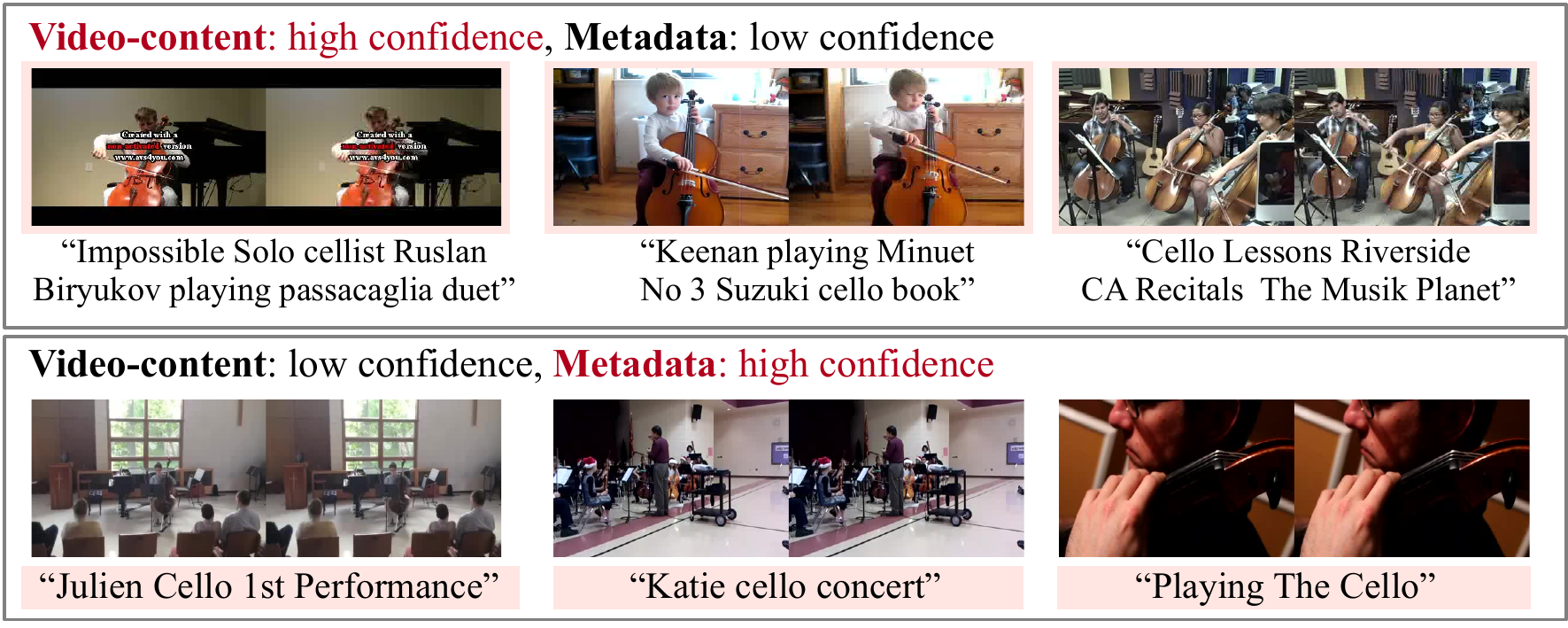}
    \label{fig:hard3_2}
}
\vspace{-4mm}
\caption{Selected samples of `cello performance' class in FCVID.
(a) Regions of selected samples are shaded: 
    \textcolor[rgb]{0.22,0.54,0.85}{max (blue)}, 
    \textcolor[rgb]{0.22,0.66,0.22}{average (green)}, and 
    \textcolor[rgb]{0.86,0.73,0.26}{product (yellow)}.
(b) Examples that are hard for only one modality. 
The upper and lower row shows examples with high confidence for metadata and concat modality, respectively 
    (upper left and lower right samples in (a), respectively).
    }
\vspace{-3mm}
\label{fig:hard3} 
\end{center} 
\end{figure*}

\subsection{Are hard samples actually selected?}
\label{sec:hard}
The motivation of MMCo is to select hard positive 
examples that cannot be selected by the previous approach.
We performed three analyses to validate MMCo really select 
\textit{hard} samples.

\textbf{Performance by difficulty:}
We evaluated the performance of example selection 
in terms of the difficulty, or easiness, of the samples.
The objective of this experiment was to show that the samples 
that cannot be detected by initial classifiers 
can be rescued by the proposed approach.
Therefore, we define easiness as a classification score 
using the classifier trained on the top 30\% of samples with 
a high pseudo label's confidence $v^0_n$.
We divided the positive samples in the training data into three sets
in terms of by easiness, 
i.e., \textit{easy}, \textit{normal}, and \textit{hard};
the samples were sorted by easiness score for each class, 
where the 30, 40, and 30th percentiles of samples are selected 
as easy, normal, and hard, respectively.
Table~\ref{tab:hard1} shows the example selection performance 
of each difficulty set.
MMCo significantly improves the performance of hard samples
in terms of both mAP (from 0.256 to \textbf{0.506}) and 
recall (0.299 to \textbf{0.579}), while there is no big difference for 
easy and normal samples.

\textbf{Distribution of difficulties of selected examples:}
We performed a more detailed analysis of selected examples.
We compare the distribution of easiness (defined above) 
with and without MMCo.
Figures~\ref{fig:dist1} and \ref{fig:dist2} shows the results 
with and without MMCo, respectively,
where the x-axis shows the rank of easiness scores among all positive-labeled 
examples, which is normalized to [0,1].
The MMCo selected positive hard examples with the scores in the range [0,0.5], 
while the method without MMCo did not select them often.
Note that there is not a big difference in the number of selected 
negative examples (i.e., false positive samples),
which shows MMCo selects only true positive examples 
from hard samples.

\textbf{Qualitative analysis:}
We qualitatively analyzed the selected examples.
We picked `cello performance' class of FCVID and 
watched the detail of the selected examples. 
We again used the classification scores of the classifier trained on 
top 30\% examples, i.e., the same condition under which samples were first
selected by MMCo.
Figure~\ref{fig:hard3_1} shows the metadata and concat classifier 
scores of positive labeled training samples, 
where true and false positive examples are depicted in red and blue.
The selected samples of each voting scheme is shaded in color
(when using only metadata and concat scores).
The results suggest why max voting performed better than other voting schemes.
We can see that the samples with high confidence for one modality are likely to be positive,
while the samples with low confidence are not always negative.
For example, 
there are some positive samples with zero confidence for one modality
(lower right and upper left in the figure),
which are hard samples for the modality but selected by using max voting.
Figure~\ref{fig:hard3_2} shows examples of these hard positives.
Metadata classifier assigns high scores to simple titles (second row)
and low scores to long titles that are difficult to parse (first row).
By learning with such hard examples,
metadata classifiers learns the \textit{key} words or phrases of 
`cello performance' such as `cellist', `playing', and `cello lessons'.
Video-content classifier (concat) is also improved by learning with
examples that are hard from the visual perspective (second row).


\begin{table*}[t] 
\begin{center}
\begin{tabular}{@{}l|c c c c c|c c c|c c c@{}}
\toprule
    & \multicolumn{5}{c}{\bf Modality in training} & \multicolumn{3}{c}{{\bf Sample selection}} & \multicolumn{3}{c}{{\bf Test performance}} \\ 
 \cmidrule(lr){2-6}\cmidrule(lr){7-9}\cmidrule(l){10-12}
{\bf Method}     & RGB  & Motion & Audio  & Concat   & Metadata & mAP      & precision& recall     & mAP      & Prec@10 & Prec@100   \\
\midrule
\multirow{5}{*}{Unimodal} 
                 & \cmark &        &        &        &        & 0.538 & 0.603 & 0.632& 0.612 & 0.835 & 0.752  \\
                 &        & \cmark &        &        &        & 0.539 & 0.601 & 0.632& 0.605 & 0.841 & 0.755 \\
                 &        &        & \cmark &        &        & 0.328 & 0.512 & 0.592& 0.594 & 0.788 & 0.658  \\
                 &        &        &        & \cmark &        & 0.640 & 0.656 & 0.645& 0.621 & 0.845 & 0.766 \\
                 &        &        &        &        & \cmark & 0.676 & 0.632 & 0.702& 0.639 & 0.868 & 0.777 \\\midrule
\multirow{6}{*}{Multimodal}                
                 & \cmark & \cmark & \cmark &        &        & 0.663 & 0.621 & 0.714 & 0.626 & 0.844 & 0.769 \\
                 &        &        &        & \cmark & \cmark & \textbf{0.781} & \textbf{0.693} & 0.801 & 0.660 & \textbf{0.880} & 0.791 \\
                 & \cmark &        &        &        & \cmark & 0.749 & 0.652 & 0.753 & 0.646 & 0.857 & 0.765 \\
                 & \cmark & \cmark &        &        & \cmark & 0.774 & 0.682 & 0.780 & 0.661 & 0.860 & \textbf{0.793} \\
                 & \cmark & \cmark & \cmark &        & \cmark & 0.772 & 0.688 & 0.783 & 0.662 & 0.873 & 0.784 \\
                 & \cmark & \cmark & \cmark & \cmark & \cmark & 0.770 & 0.694 & \textbf{0.803} & \textbf{0.663} & 0.862 & 0.781 \\
\bottomrule
\end{tabular}
\end{center}
\caption{Combinations of modalities used in training. The testing classifier is the same (concat) for all settings.}
\vspace{-5mm}
\label{tab:modality} 
\end{table*}

\subsection{Which modalities improve performance?}
To investigate which modalities improve performance,
we tested with varied combination of modalities.
Since our main interest is the improvement in example selection performance,
we only changed training modalities  
that are used to compute scores for example selection.
We used the concat feature in the testing in all experiments.
Table~\ref{tab:modality} shows the results for various modality
combinations;
rows 1--5 are for only a single modality,
while rows 6--11 are for multiple modalities.
The results provide several important findings.
1) The results using both video-content and metadata modality (rows 7--11)
are much better than those using only either type of features (rows 1--6).
This suggests that using modalities with complementary information is 
more important than using just multiple modalities.
2) Among the video-content related features, 
P3D contributed significantly to performance improvements,
while audio did not contribute very much (row 8--10).
3) The results for Concat + Metadata (row 7) are comparable 
to those of using more than two classifiers,
which is a practically reasonable choice in terms of 
the trade-off between the training cost and test accuracy.

%
%

\begin{figure}[t] 
\begin{center}
\includegraphics[width=1.00\linewidth]{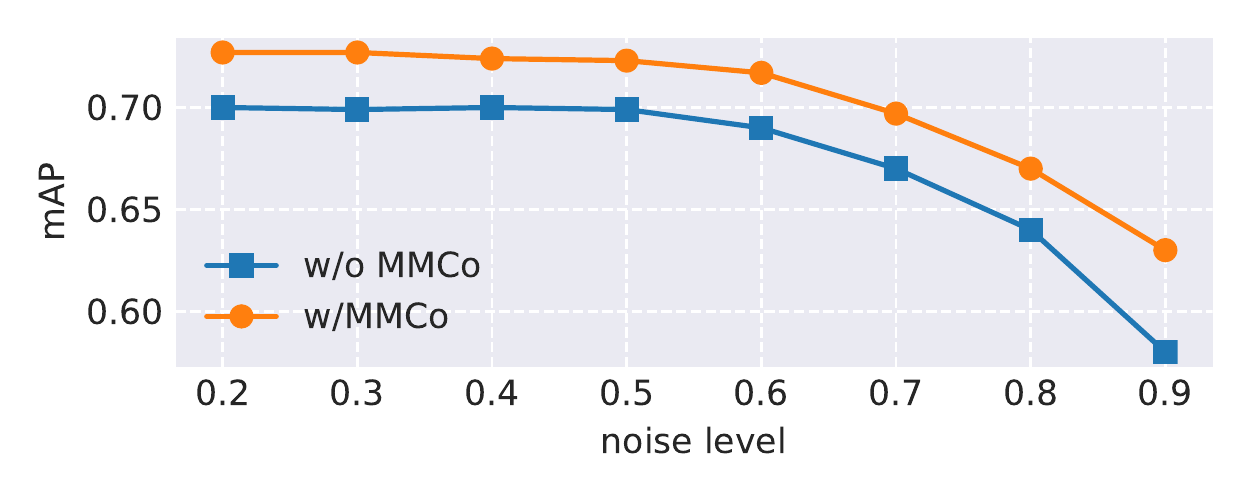}
\vspace{-8mm}
\caption{Noise level vs test mAP on FCVID.}
\vspace{-4mm}
\label{fig:noise2}
\end{center} 
\end{figure}

\begin{figure*}[t] 
\begin{center}
\includegraphics[width=1.00\linewidth]{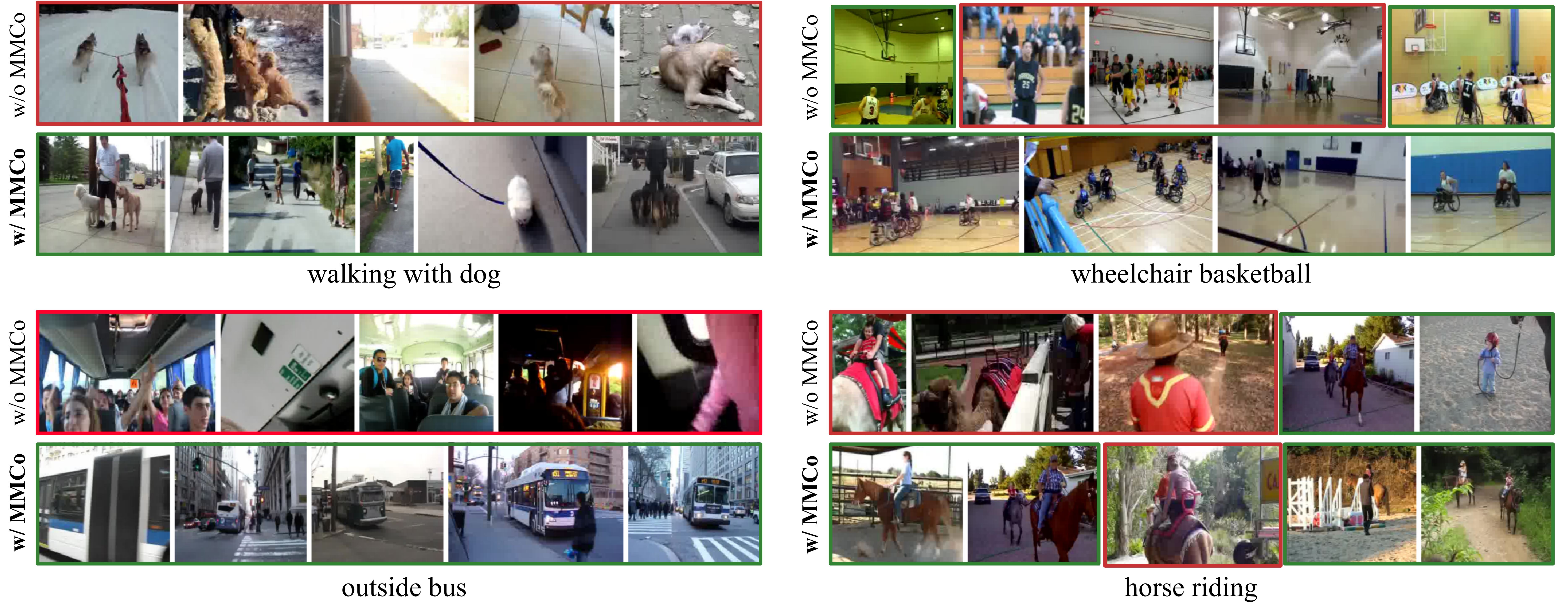}
\vspace{-7mm}
\caption{Top retrievals on FCVID dataset. Retrievals were performed by 
using the classifier trained with the webly label,
where the results are sorted by classification score and marked
as \textcolor[rgb]{0.8, 0.2, 0.2}{negative (red)} 
or \textcolor[rgb]{0.2, 0.5, 0.2}{positive (green)}.
}
\label{fig:qual} 
\vspace{-3mm}
\end{center} 
\end{figure*}

\subsection{Noise robustness}
We evaluated the robustness of our approach to noise
by manually controlling the amount of noise in the training data.
Since the training data of our task is noisy,
there are true and false positives among the positive-labeled samples.
We varied the noise level $n_{FP}/(n_{TP}+n_{FP})$ in the range of 
0.2 to 0.9, where $n_{TP}$ and $n_{FP}$ are the number of 
true and false positives, respectively.
We fixed the true positive samples and randomly selected 
false positive samples as artificial noisy samples.
Figure~\ref{fig:noise2} compares the performances of our approach with 
and without MMCo on FCVID for different noise levels.
Our approach (w/ MMCo) was consistently better than the baseline 
(w/o MMCo) with any amount of noise.
It scored 0.63 in mAP at a noise level of 0.9, 
which demonstrates that it can work on a very noisy dataset.

\subsection{Qualitative examples}
Figure~\ref{fig:qual} shows the top retrieval results on FCVID
by using the classifiers trained by webly labeled learning.
The results with and without MMCo are compared
to qualitatively demonstrate the effectiveness of our approach.
While the approach without MMCo detects many false positives
of very similar concept, the approach with MMCo rejects such false
positives by effectively using the metadata information in training.
Since our trained classifier is a simple linear classifier, 
this retrievals can be performed in less than one second 
from 48,435 videos.
It is also easy to extend million or billion-scale retrievals 
by using the techniques of approximate nearest neighbor 
search~\cite{Jegou2011}.


\section{Conclusion} 
We presented a multimodal co-training (MMCo) algorithm, 
a simple and effective
method for learning concept classifier from videos on the web.
MMCo can selects good examples from noisy training data by 
jointly learning classifiers of multiple modalities
and selecting examples based on their consensus.
We conducted an extensive experimental analysis to demonstrate 
the effectiveness of MMCo,
which demonstrated better example selection and consistent improvements
in webly labeled learning on FCVID dataset over baseline methods.
Our detailed analysis from multiple aspects further 
examined the behavior of MMCo and provided several useful insights 
about how to exploit multimodal information effectively in webly supervised learning.

\bibliographystyle{ACM-Reference-Format}
\balance
\bibliography{Webly,add} 

\end{document}